\documentclass[aps,prd,groupedaddress,showpacs,twocolumn]{revtex4}
\usepackage{amsmath}
\usepackage[parfill]{parskip}    
\usepackage{float}
\usepackage{caption}
\usepackage{subcaption}
\usepackage{graphicx}
\usepackage{bbm}
\usepackage{graphicx}
\usepackage{amssymb}
\usepackage{epstopdf}
\usepackage{color}
\usepackage{ulem}
\usepackage{graphicx}
\usepackage{pdfpages}
\usepackage[colorlinks=true,citecolor=blue,urlcolor=magenta,breaklinks]{hyperref}

\begin{document}

\title{\bf Thermodynamics of scale-dependent Friedmann equations}

\author{Pedro \surname{Bargueño}}
\email{pedro.bargueno@ua.es}
\affiliation{Departamento de F\'{i}sica Aplicada, Universidad de Alicante, Campus de San Vicente del Raspeig, E-03690 Alicante, Spain.}

\author{Ernesto Contreras}
\email{econtreras@usfq.edu.ec}
\affiliation{Departamento de F\'isica, Colegio de Ciencias e Ingenier\'ia, Universidad San Francisco de Quito,  Quito, Ecuador.}

\author{\'Angel Rinc\'on}
\email{angel.rincon@pucv.cl}
\affiliation{Instituto de F{\'i}sica, Pontificia Universidad Cat{\'o}lica de Valpara{\'i}so, Avenida Brasil 2950, Casilla 4059, Valpara{\'i}so, Chile.}

\begin{abstract}

In this work, the role of a time-varying Newton constant under the scale-dependent approach is investigated in the thermodynamics of the Friedman equations. In particular, we show that the extended Friedman equations can be derived either from equilibrium thermodynamics when the non-matter energy momentum tensor is interpreted as a fluid or from non-equilibrium thermodynamics when an entropy production term, which depends on the time-varying Newton constant, is included. Finally, a comparison between black hole and cosmological thermodynamics in the framework of  scale--dependent gravity is briefly discussed.

\end{abstract}

\maketitle

\section{Introduction}
\label{Intro}
It is well known that the “standard model” of the Universe is based on the Cosmological Principle \cite{Liddle:1998ew}. The latter states that our Universe is homogeneous and isotropic when viewed on large scales (i.e., $\ge$ 100 Mpc, spatial scales). Taking into account the two above facts, the FLRW metric emerges \cite{Ellis:1998ct}. Thus, we can use the Einstein’s field equations for the perfect fluid to obtain the Friedman equations. Including the cosmological constant, the equations look like
\begin{eqnarray}
        \label{ff1}
	H^2+\frac{\kappa}{a^2}&=&\frac{\Lambda}{3}+\frac{8 \pi G}{3}\rho_{s}\\
	H'+H^2&=& \frac{\Lambda}{3}-\frac{4 \pi G}{3}\left(\rho_{s}+ 3 p_{s} \right),
        \label{ff2}
\end{eqnarray}
where a concrete relation between $\rho_s$ and $p_s$, an equation-of-state (EoS hereafter), is always required. 

It should be pointed out that $\Lambda$CDM (one of the most basic FLRW models) assumes that dark energy is a cosmological constant $\Lambda$.
This model is able to explain many of the current observations, although certain evidence suggests that such a model is inadequate to explain some features of cosmic evolution and the large scale structure \cite{Melia:2014vva}.

In general, we have a huge variety of cosmological models (see \cite{Ellis:1998ct} and references therein). Originally, they were based on general relativity but, in light of the above-mentioned discrepancies, the community has moved to study cosmological problems taking advantage of alternative models of gravity. Avoiding unnecessary details, we should mention a few approaches used up to now, for example: scalar-tensor theories \cite{Brans:1961sx,Bergmann:1968ve,Faraoni:2004pi}, vector-tensor \cite{Hellings:1973zz,Will:1972zz} and just tensor theories as Starobinsky \cite{Starobinsky:1979ty,Starobinsky:1980te}, Gauss–Bonnet, $f(R)$ \cite{DeFelice:2010aj} and Lovelock gravity \cite{Lovelock:1971yv}. In addition, a recent and novel approach can be included into the subset of alternatives theories of gravity; the so--called  scale-dependent gravity, which basically extend the classical solutions by the inclusion of scale-dependent couplings (see \cite{SD1,SD0,SD2,SD3,SD4,SD5,SD6,SD7,SD8,SD9,SD10,SD11,SD12,SD13,SD14,SD15,astro1,astro2,cosmo1,cosmo2,Contreras:2018gpl}
and references therein).

Regardless of the approach used to make progress, the relevance of thermodynamics properties in the cosmological context is doubtless. In black hole physics, for example, it is well-known that the temperature is proportional to its surface gravity (at the black hole horizon), and the entropy is proportional to its horizon area \cite{Hawking:1974sw,Majumdar:1998xv}.
Such quantities, and the corresponding black hole mass, obey the first law of thermodynamics \cite{Carlip:2014pma}.
Along the years, a transparent relation between the Einstein's field equations and the black hole thermodynamics has been investigated in detail. In the cosmological context, however, some progress has been made \cite{Cai:2005ra,CaiCao2007}, but still more research is required. 

Thus, thermodynamic properties are always a suitable ingredient to be included and properly analyzed. Clearly, black hole physics is usually a substantial motivation to tries to understand how the thermodynamics can serves to gets the Einstein's field equations in a cosmological context.
In particular, as has been pointed out in Refs. \cite{Cai:2005ra,CaiCao2007}, taking advantage of the first law of thermodynamics to the apparent horizon of a Friedmann-Robertson-Walker universe and considering the entropy given by a quarter of the apparent horizon area, it is possible to derive the Friedmann equations describing the dynamics of the universe with any spatial curvature. Following the same spirit, we will deal with the same problem, but now in an alternative formalism: the scale-dependent scenario.

This work is organized as follows, after this brief introduction, we will discuss in section II, the main features of scale-dependent gravity as well as the basic equations to be considered. Then, in Section III the corresponding generalized Einstein's field equations in cosmology are derived. After that, in the next section the thermodynamics considerations of these models are discussed. Finally, some concluding remarks are made in the last section.

\section{Scale-dependent gravity}
\label{SDgravity}

This sections aims to summarize the main ingredients of the scale-dependent formalism. Up to now, the idea has been significantly implemented in several contexts, viz, alternative black hole solutions at lower and higher dimensions, 
relativistic stars and also in cosmology (see \cite{SD1,SD0,SD2,SD3,SD4,SD5,SD6,SD7,SD8,SD9,SD10,SD11,SD12,SD13,SD14,SD15,astro1,astro2,cosmo1,cosmo2,Contreras:2018gpl} for details). In addition, certain black hole properties has been reviewed in light of the above-mentioned formalism, for instance, the computation of quasinormal modes of black holes for certain geometries.

Scale-dependent gravity takes advantage of the seminal work of Reuter and Weyer (see \cite{Reuter2004} for details) which has served as an inspiration for a subsequent set of works (see for instance \cite{Koch2015} and also \cite{Contreras:2013hua}). 

A detailed discussion can be found in Ref. \cite{Reuter2004,Bonanno:2000ep,Bonanno:2020qfu}, but the basic idea is to promote the couplings, which appears in the gravitational action, to scale-- quantities. It should be mentioned that such approach is also quite common in high energy theories.

As usual, the starting point will be a general action, namely, a scale--dependent version of the Einstein-Hilbert action, without ghosts or any other contributions given by 
\begin{eqnarray}\label{action}
S[g_{\mu\nu},k]=\int \mathrm{d}^{n}x\sqrt{-g}
\bigg[\frac{1}{2 \kappa_{k}} \bigg(R-2\Lambda_{k}\bigg)\bigg],
\end{eqnarray}
\\
\\
where $k$ is a scale--dependent field considered as a renormalization scale and
$\kappa_{k} \equiv 8 \pi G_{k}$ is the Einstein coupling. Additionally, we have two more couplings, i.e., 
$G_{k}$ which represent the scale-dependent gravitational coupling and $\Lambda_{k}$ that corresponds to the scale-dependent cosmological couplings.
Now, we compute the corresponding modified Einstein's equations taking variations with respect to the metric field $g_{\mu\nu}$ to obtain
\\
\begin{eqnarray}\label{einstein}
G_{\mu\nu}+g_{\mu\nu}\Lambda_{k}=-\Delta t_{\mu\nu},
\end{eqnarray}
\\
being the additional tensor $\Delta t_{\mu\nu}$ defined as
\\
\begin{eqnarray}\label{nme}
\Delta t_{\mu\nu}=G_{k}\Bigl(g_{\mu\nu}\square -\nabla_{\mu}\nabla_{\nu}\Bigl)G_{k}^{-1}.
\end{eqnarray}
\\
Then, to supplement our set of equations, we take the variation of the effective action with respect to the scale field $k(x)$, i.e., 
\begin{equation}\label{scale}
\frac{\mathrm{d}}{\mathrm{d} k} S[g_{\mu \nu}, k] =0,
\end{equation}
considered as an a posteriori condition towards background independence. 

In practice, both Eqs. (\ref{einstein}) and (\ref{scale}) are sufficient to close the system. However, given that the $\beta$-functions are in general unknown, it could be a more convenient approach to consider that both couplings, $G_{k}$ and $\Lambda_{k}$, are promoted to be 
space--time dependent function, depending of the scale field, $k(x)$, so that the corresponding couplings are written as $G(x)$ and $\Lambda(x)$. 
In this sense, the problem is, basically, looking for the solution of the metric potentials, the Newton and cosmological function. 
Following the above mentioned idea and considering an adequate  line element, we will be able to solve the problem in situations with a high symmetry.
\\
\\

With these ideas in mind, scale--dependent gravity can be considered, in some sense, as a special kind of scalar-tensor
theory, although the improved action is not supposed to be varied neither with respect to $G(x)$ nor to $\Lambda(x)$ \cite{Reuter2004}. 
The total action we consider is
\\
\begin{equation}
	\label{pp}
	S_{\mathrm{T}}=\int d^{4}x \sqrt{-g} \bigg(\frac{R}{16 \pi G(x)}-\frac{\Lambda(x)}{8\pi G(x)} \bigg)+S_{\mathrm{mat}},
\end{equation}
\\
where $S_{\mathrm{mat}}$ encodes the matter sector of the theory.
Then, after varying Eq. (\ref{pp}) with respect to the metric, we arrive at
\begin{equation}
\label{gen}
	G_{\alpha \beta}+G\Big(g_{\alpha \beta}\square-\nabla_{\alpha}\nabla_{\beta} \Big)G^{-1}+\Lambda g_{\alpha \beta}=
	8 \pi G T_{\alpha \beta} 
\end{equation}
\\
or
\begin{equation}
\label{gen2}
	\mathcal{S}_{\alpha\beta}\equiv G_{\alpha \beta}+\Delta t_{\alpha \beta}+\Lambda g_{\alpha \beta}-
        8 \pi G T_{\alpha \beta}=0,
\end{equation}
where, as stated before, $G$ and $\Lambda$ have to be understood as functions of the space--time points. Be aware and notice that a common redefinition of the energy-momentum tensor can be used, which accounts for the running of Newton's coupling via the tensor $\Delta t_{\mu \nu}$. Thus, we may think that the inclusion of a running Newton's coupling is a natural mechanism to obtain an effective energy-momentum tensor in more generalized scenarios.

\section{Generalized Einstein equations in a cosmological setting}
\label{Calculations}


Let us consider a line element written as
\\
\begin{equation}
	\label{metric}
	ds^2= -dt^2+a(t)^{2}\bigg(\frac{dr^2}{1-\kappa r^2}+d\Omega^2\bigg),
\end{equation}
\\
where $d\Omega^2$ is the line element four the round 2-sphere, $a(t)$ is the scale factor and $\kappa$ stands for the spatial curvature.
Even more, within a time--dependent context, $\Lambda$ and $G$ turn into functions of the time coordinate only. In this particular case, the 
generalized (vacuum) equations are written as:
\\
\begin{widetext}
\begin{eqnarray}
	\label{eq1}
	-\frac{3 a'(t) G'(t)}{a(t) G(t)}+\frac{3 \left(a'(t)^2+\kappa \right)}{a(t)^2}-\Lambda (t)&=0& \\
	\label{eq2}
	 -a(t) G(t) \left(2 a'(t) G'(t)+a(t) G''(t)\right)+G(t)^2 \left(2 a(t) a''(t)+a'(t)^2+a(t)^2 (-\Lambda (t))+\kappa \right)+2 a(t)^2 G'(t)^2&=0&.
\end{eqnarray}
\end{widetext}

Note that we have two equations for three unknowns, $a(t),G(t),\Lambda(t)$; {\it i.e.}, the system is undetermined. For simplicity, we can consider the case $\Lambda(t)=0$ which leads to following conclusions: 
\\
\\
(i) Eq. (\ref{eq1}) can be solved for $G(t)$, obtaining
\\
\begin{equation}
	G(t)=C_{1}\exp \left(\int_1^t \frac{a'(y)^2+\kappa }{a(y) a'(y)} \, d y\right).
\end{equation}
\\
\\
(ii) In this case, Eq. (\ref{eq2}) turns into
\\
\begin{equation}
\frac{\left(a'(t)^2+\kappa \right) \left(a(t) a''(t)+a'(t)^2+\kappa \right)}{\left(\kappa  r^2-1\right) a'(t)^2}=0.
\end{equation}
\\
\\
(iii) We observe that only in the $\kappa=0$ case the solution for $a(t)$ is unique.
\\
\\
(iv) The idea is to solve Eqs. (\ref{eq1}) and (\ref{eq2}) in such a way that one is able
to identify two constants of integration, $G_{0}$ (the non-running Newton constant) and $t_{\epsilon}$,
the running time scale which signals when scale-dependent effects are supposed to appear.
\\
\\
(v) When $t_{\epsilon}\rightarrow \infty$, 
one should recover the corresponding general relativistic solution ($\Delta t_{\alpha \beta}=0$) which, given the FRW ansantz we have chosen together with $\Lambda(t)=0$, it should be the Minkowski space--time. This implies that we are looking for an improved 
(in the scale-dependent sense), homogeneous and isotropic space--time emerging from the Minkowski background when scale-dependence is switched on.

\subsection{A scale-dependent radiation universe}

The following functions solve Ecs. (\ref{eq1}) and (\ref{eq2}) in the case of interest:
\\
\begin{eqnarray}
	G(t)&=&G_{0} \sqrt{1+\frac{2t}{t_{\epsilon}}}\nonumber \\
	a(t)&=&a_{0}\sqrt{1+\frac{2t}{t_\epsilon}}\label{at}, 
\end{eqnarray}
\\
where $t_{\epsilon}>0$ for physically relevant solutions. Interestingly, Eq. (\ref{at}) resembles the profile of the scale--factor obtained in the framework of loop quantum cosmology in the sense that the singularity at $t=0$ has been removed (see \cite{zhu}, for example). Of course, this is a formal relationship. In loop quantum cosmology the power law is 1/6 instead 1/2 and the critical time $t_{\epsilon}$ should be replaced by the Planck time.\\

In principle, in view of Eqs. (\ref{gen2}), one can assign some kind of energy densities and pressures associated with
the non-matter energy-momentum tensor, by interpreting it as some kind of {\it curvature fluid}, as is usually established
in $f(R)$ theories \cite{Capozziello2011}. In our case of interest, we can define this effective energy-momentum tensor as
\\
\begin{equation}
\mathcal{T}_{\alpha \beta}=-8 \pi G(t) \Delta t_{\alpha \beta}.
\end{equation}
\\
Interestingly, within the fluid interpretation we obtain the corresponding density and pressure as
\\
\begin{eqnarray}
	\rho_{s}(t) = -g^{tt}\mathcal{T}_{tt} &=& \frac{3}{8 \pi  G_{0} t_{s}^2 \left(\frac{2 t}{t_{s}}+1\right)^{5/2}} \\
	p_{s}(t) = -g^{rr}\mathcal{T}_{rr}&=& \frac{1}{8 \pi  G_{0} t_{s}^2 \left(\frac{2 t}{t_{s}}+1\right)^{5/2}},
\end{eqnarray}
\\
which satisfy the radiation condition,
\\
\begin{equation}
-\rho_{s}+ 3 p_{s} = 0. \label{radia}
\end{equation}
\\
We note that $a(t)=a_{0}$, $G(t)=G_{0}$, $\rho_{\epsilon}=p_{\epsilon}=0$  when $t_{\epsilon}\rightarrow \infty$, as required to reach
the Minkowskian limit. Note that, although
Eq. (\ref{radia}) could lead to conclude that our findings serve to model a radiation dominated era, it should be emphasized that our solution is a vacuum one and (\ref{radia}) arises as a consequence of the dependence of the Newton coupling as a function of the space--time points (encoded in the non--matter energy momentum tensor). Even more, there are not reasons to believe that $t_{\epsilon}$ allows to connect different epochs.  

Interestingly, the curvature scalars are given by
\\
\begin{eqnarray}
	R = g^{\alpha \beta}R_{\alpha \beta}&=&0 \\
	K =R^{\alpha\beta\gamma\delta}R_{\alpha\beta\gamma\delta}&=&\frac{24}{\big(2 t+ t_{\epsilon}\big)^4}\\
	\mathrm{Ric}^2=R^{\alpha \beta} R_{\alpha \beta}&=& \frac{12}{\big(2 t+ t_{\epsilon}\big)^4}\\
\end{eqnarray}
\\
showing that the corresponding solution is regular everywhere (remember we are considering $t_{\epsilon}>0$).

\section{Thermodynamical considerations}
Let us write Eqs. (\ref{eq1}) and (\ref{eq2}) in a Friedmann-like form. By introducing the usual Hubble parameter $H=\dot a/a$, we arrive to
\\
\begin{eqnarray}
	\label{f1}
	H^2+\frac{\kappa}{a^2}&=&\frac{\Lambda}{3}+H\frac{G'}{G}\\
	H'+H^2&=& \frac{\Lambda}{3}+\frac{1}{G}\left(H G'+ G'' -\frac{2\, (G')^2}{G}\right)
	\label{f2}.
\end{eqnarray}
\\
\\
As commented before, we can define an effective scale-dependent energy-momentum tensor, $\mathcal{T}_{\alpha \beta}$, 
from which the corresponding energy density and
pressure are given by
\\
\begin{eqnarray}
	\rho_{s}&=&\frac{3}{8 \pi}\frac{H G'}{G^2}\\
	p_{s}&=& -\frac{1}{4\pi G^2}\left(\frac{3 H G'}{2}+G''-2\left(\frac{G'}{G}\right)^2 \right),
\end{eqnarray}
\\
\\
such that Eqs. (\ref{f1}) and (\ref{f2}) acquire the usual form:
\\
\begin{eqnarray}
        \label{ff1}
	H^2+\frac{\kappa}{a^2}&=&\frac{\Lambda}{3}+\frac{8 \pi G}{3}\rho_{s}\\
	H'+H^2&=& \frac{\Lambda}{3}-\frac{4 \pi G}{3}\left(\rho_{s}+ 3 p_{s} \right),
        \label{ff2}
\end{eqnarray}
\\
\\
where $\Lambda$ and $G$ have to be understood as $\Lambda(t)$ and $G(t)$, respectively.
\\
\\
Finally, we note that, when a matter sector is considered, the corresponding energy-momentum tensor contributes to Eqs. (\ref{ff1}) and
(\ref{ff2}) with its own densities and pressures, $\rho_{m}$ and $p_{m}$, such that the structure of the previously mentioned
equations remains the same with the sources being $\rho_{m}+\rho_{s}$ and $p_{m}+p_{s}$.
\\
\\
At this point, several comments are in order. First, Cai and Kim \cite{Cai:2005ra} proved that it is possible to
derive the Friedmann equations describing the dynamics of the
universe with any spatial curvature by 
applying the first law of thermodynamics to the apparent horizon of a FRW geometry, and considering that the entropy is given by 
a quarter of the
apparent horizon area. 
And second, by using the tunneling approach of Parikh and Wilczek 
\cite{Parikh2000,Parikh2002}, Cai {\it et al.} showed \cite{Cai2009} that there exists Hawking radiation for a locally
defined apparent horizon of a FRW universe with any spatial curvature. In essence,
the authors of Ref. \cite{Cai2009} shown that the inverse temperature is given by
\\
\begin{equation}
\label{hor}
	\beta= 2\pi \tilde r_{a},
\end{equation}
\\
where $\tilde r_{a}=H^{-1}$ is the location of the apparent horizon of a flat ($\kappa=0$) FRW universe.
\\
In our case of interest, it is straightforward to show that the scale-dependent version of the emission appears at a inverse 
temperature given by
\begin{equation}
\label{temp}
	\beta_{s}=2\pi\big(t_{\epsilon}+2t \big),
\end{equation}
which diverges for $t_{\epsilon}\rightarrow \infty$ (Minkowskian limit).
\\
\\
This fact supports a first evidence of an existing relationship between thermodynamics and the scale--dependent Friedmann equations
given by Eq. (\ref{f1}) and (\ref{f2}). In the following lines we will show that Cai and Kim approach of Ref. \cite{Cai:2005ra} can be extended 
to the scale dependent case under appropriate assumptions.
\\
\subsection{Scale-dependent Friedmann equations from thermodynamics}

In this section we will follow closely the techniques employed in Ref. \cite{Cai:2005ra}, extending them when necessary.
\\
\\
First of all, let us write Eq. (\ref{metric}) as
\\
\\
\begin{equation}
	ds^2= h_{a b}dx^a dx^b + \tilde r^2 d\Omega^2,
\end{equation}
\\
\\
where $\tilde r = a(t)r$, $x^0=t$, $x^1=r$ and $h_{ab}=\mathrm{diag}\left(-1,\frac{a^2}{1- \kappa r^2} \right)$. The dynamical apparent,
$\tilde r_{a}$
horizon is determined by the relation $h^{ab}\partial_{a}\tilde r\partial_{b}\tilde r=0$ which, in this case, is given by
\\
\\
\begin{equation}
	\tilde r_{a}=\left(H^2+\frac{\kappa}{a^2}\right)^{-1/2}.
\end{equation}
\\
\\
This dynamical apparent horizon has been argued to be a causal horizon with both gravitational entropy and surface gravity 
\cite{Hayward1999,Bak2000}. Therefore, we will employ it in the first law of thermodynamics in order to deduce the generalized
Friedmann equations.
\\
\\
After defining the work density at $\tilde r_{a}$, $W$, as the work done by a change
of the apparent horizon,
\\
\\
\begin{equation}
	W=-\frac{1}{2}T^{a b}h_{a b},
\end{equation}
\\
\\
and the energy-supply vector at $\tilde r_{a}$, $\Psi_{a}$, as the total energy flow
through it \cite{Hayward1999,Bak2000} as
\\
\\
\begin{equation}
	\Psi_{a}=T_{a}^{b}\partial_{b}\tilde r + W \partial_{a}\tilde r,
\end{equation}
\\
\\
we arrive to Hayward's unified first law \cite{Hayward1999},
\\
\\
\begin{equation}
	\label{Hay1}
\partial_{a} E = A \Psi_{a} + W \partial_{a} V.
\end{equation}
\\
\\
In Eq. (\ref{Hay1}), $A=4 \pi \tilde r^2$, $V=\frac{4}{3}\pi \tilde r^3$ and the total energy inside the spherical space bounded
by $\tilde r$ is
\\
\\
\begin{equation}
	E=\frac{\tilde r}{2 G}\left(1-h^{ab}\partial_{a}\tilde r\partial_{b}\tilde r \right).
\end{equation}
\\
\\
At this point it is important to note that both the work density and the energy flow include a contribution
coming from the non--matter energy momentum tensor, $\Delta t_{a b}$. In addition, the total energy depends on the time--dependent Newton's
constant, $G=G(t)$.
This issue was raised by Cai and Cao \cite{CaiCao2007}, who pointed out that ``an interesting question is whether the field equations for non-Einstein gravity
can be written to a form as the unified first law in Einstein
gravity". In the same spirit, after rewriting the field equations for scale-dependent gravity in the form of the Einstein gravity by introducing an effective energy-momentum tensor, we assume that the unified first law can
be safely employed.

Even more, the heat flow is related to the change of energy of the given system. Therefore, the entropy is associated with the energy-supply term, which can
be rewritten, in our case, as
\begin{equation}
\label{APsi}
    A \Psi = \frac{\kappa}{8 \pi} \nabla \left(\frac{A}{G} \right)+\frac{\tilde r}{G}\nabla\left(\frac{E\, G}{\tilde r}\right),
\end{equation}
\\
where $\kappa$ is the surface gravity defined as
\begin{equation}
\label{kappa}
        \kappa = \frac{1}{2\sqrt{-h}}\partial_{a} \left(\sqrt{-h}h^{ab}\partial_{b}\tilde r \right).
\end{equation}
\\
Note that the main difference with the case studied in Ref. \cite{Cai:2005ra} is the inclusion of $G$ inside the differential terms. Even more, as the second term of the rhs of Eq. (\ref{APsi}) vanishes on the apparent horizon, we are left with an entropy and a temperature
given by
\\
\begin{eqnarray}
\label{S}
        S&=& \frac{A\left(\tilde r_{a}\right)}{4 G\left(\tilde r_{a} \right)} \\
        T&=&\frac{1}{2 \pi \tilde r_{a}}
        \label{T}.
\end{eqnarray}
At this point a couple of comments are in order: (i) the apparent horizon and its corresponding temperature are modified following Eqs. (\ref{hor}) and (\ref{temp}), respectively; (ii) the entropy acquires a form similar to that obtained in Brans-Dicke and
$f(R)$ theories (see, for example, Ref. \cite{BarguenoUA2020} and references therein) and (iii), the general relativistic limit (which in this case collapses to the Minkowskian limit) is obtained when $t_{\epsilon}\rightarrow \infty$. 
\\
The heat flow through the apparent horizon is the change of the energy inside it. Moreover, assuming a perfect fluid energy-momentum tensor for the matter sector, we find that \cite{Cai:2005ra}
\begin{equation}
    -d E = - A \Psi = A \left(\rho+p \right) H \tilde r_{a} dt.
\end{equation}
Finally, if the Clasius law,
\begin{equation}
-d E = T dS    
\end{equation}
is applied with $T$ and $S$ given by 
Eqs. (\ref{S}) and (\ref{T}), it is to show that it is not possible to arrive to the scale-dependent Friedmann equations. Interestingly, a similar no-go behaviour has been found in scalar-tensor and $f(R)$ theories \cite{Akbar2006,CaiCao2007}.
\\
\\
There is a simple way of circumventing this problem. First, note that the temperature is a purely geometric quantity and, therefore, it does not depend on the specific theory one is dealing with. Second, instead of taking a perfect fluid energy-momentum for matter, $T_{\alpha \beta}$, let
us consider $T_{\alpha \beta}+\mathcal{T}_{\alpha \beta}=T_{\alpha \beta}-8 \pi G(t) \Delta t_{\alpha \beta}$. And third, let us 
take the corresponding entropy as $S=\frac{A}{4 G_{0}}$ (note that $G_{0}$ has been employed instead of $G(\tilde r_{a})$. Then, as it should be clear from Eqs. (\ref{ff1}) and (\ref{ff2}) together with the associated discussion, the procedure works out and
the corresponding scale-dependent Friedmann equations can be obtained from thermodynamics, as previously pointed out for scalar-tensor \cite{CaiCao2007} and $f(R)$ theories \cite{Akbar2006}. Finally, it has been shown \cite{PRD2014} that
the Clasius law holds when an entropy production term, $d S_{p}$, is added to it. Specifically, this entropy production term is given, in our context, by Eq. (41) of Ref \cite{PRD2014} which states that
\begin{equation}
    d S_{p}=2 \pi \tilde r_{a}^2 \frac{\dot G(\tilde r_{a})}{G^2(\tilde r_{a})}dt.
\end{equation}

\section{Final remarks and conclusions}
\label{discussion}
We close this work by briefly comparing black hole and cosmological thermodynamics in the framework of SD gravity. In the former case, the authors of Ref. \cite{BarguenoUA2020} obtained appropriate generalizations for both the entropy and energy of SD black holes, showing that no entropy production terms are needed in order to describe the theory, in complete agreement with both scalar–tensor \cite{CaiCao2007} and $f(R)$ theories \cite{Wu2008}. In the latter case, although entropy production can be introduced in order for the Clasius law to be valid, a redefinition of the energy associated with the apparent horizon can also be performed by introducing a non-equilibrium energy dissipation term \cite{PRD2014}. Therefore, non-equilibrium thermodynamics is needed in SD gravity in order to completely derive the extended Friedmann equations.
\\
\section*{ACKNOWLEDGEMENTS}
P. B. dedicates this work to Ana\'{\i}s, Luc\'{\i}a, In\'es and Ana for continuous support.
P. B. is funded by the Beatriz Galindo contract BEAGAL 18/00207 (Spain). 
The author A.~R. acknowledges DI-VRIEA for financial support through Proyecto Postdoctorado 2019 VRIEA-PUCV. Ángel Rincón dedicates this work to Elena Guilarte, who always gave him her unconditional love and support. ``I will see you again".
\\
\\

\end{document}